\documentclass[twocolumn,showpacs,aps,floatfix,prd]{revtex4}

\usepackage[normalem]{ulem}

\usepackage{graphicx}
\usepackage{dcolumn}
\usepackage{amsmath}
\usepackage{epsfig}
\usepackage{hhline}
\usepackage{soul}
\usepackage{bm}
\usepackage{tabularx,pifont,multirow}

\usepackage{longtable}
\usepackage{psfrag}

\input pubboard/babarsym

\newcommand{\ij}{{_i^j}}
\newcommand{\bn}[1]{b^n_{{#1}}}
\newcommand{\sn}[1]{s^n_{{#1}}}
\newcommand{\bw}[1]{b^w_{{#1}}}

\newcommand{\sw}[1]{s^w_{{#1}}}
\newcommand{\fn}[1]{f^n_{{#1}}}
\newcommand{\fo}[1]{f^o_{{#1}}}

\newcommand{\lifetime}[1]{\tau_{{#1}}}
\newcommand{\dm}[1]{\Delta m_{{#1}}}
\newcommand{\ot}[1]{\omega_{{#1}}}
\newcommand{\dmistag}[1]{\Delta \omega_{{#1}}}

\def\bnj#1#2{b^n_{{#1}}{^{#2}}}
\def\snj#1#2{s^n_{{#1}}{^{#2}}}
\def\bwj#1#2{b^w_{{#1}}{^{#2}}}

\def\swj#1#2{s^w_{{#1}}{^{#2}}}
\def\fnj#1#2{f^n_{{#1}}{^{#2}}}
\def\foj#1#2{f^o_{{#1}}{^{#2}}}

\def\beq{\begin{equation}}
\def\eeq{\end{equation}}
\def\bea{\begin{eqnarray}}
\def\eea{\end{eqnarray}}
\def\resp#1{} 

\def\sinphi{\sin(2\beta + \gamma)}

\def\mmiss{m_{\rm miss}}

\def\Dstarp{\Dstar{^+}}
\def\Dstarm{\Dstar{^-}}

\def\btoc{b \rightarrow c \bar u  d}
\def\btou{b \rightarrow u \bar c  d}

\def\btodstpipm{B \rightarrow \Dstarmp\pi^\pm}
\def\btodstpi{\Bz \rightarrow \Dstarp\pi^-}

\def\btodstrhopm{B\rightarrow \Dstarmp\rho^\pm}

\def\dt{\Delta t}
\def\dtErr{\sigma_{\dt}}
\def\dz{\Delta z}

\def\dttrue{\dt_{\rm tr}}

\def\mc{MC}

\def\Brec{B_{\rm rec}}
\def\Btag{B_{\rm tag}}
\def\zrec{z_{\rm rec}}
\def\ztag{z_{\rm tag}}

\def\stag{s_{_{\rm t}}}
\def\smix{s_{_{\rm m}}}

\def\dstpi{{\Dstar\pi}}
\def\dstrho{{\Dstar\rho}}

\def\comb{{\rm comb}}
\def\peak{{\rm peak}}

\def\cont{{q\overline q}}

\def\P{{\cal P}}  
\def\T{{\cal T}}  
\def\M{{\cal M}}  
\def\F{{\cal F}}  
\def\R{{\cal R}}  
\def\A{{\cal A}}  
\def\BG{\hat {\cal G}}
\def\G{{\cal G}}  

\def\r{{r^*}}
\def\rmeas{r^{*\rm meas}}
\def\rp{r'}
\def\rsq{r^{*2}}
\def\deltaPhaseP{\delta'}
\def\deltaPhase{\delta^*}

\def\dir{{\rm dir}}
\def\cas{{\rm cas}}
\def\mis{{\rm miss}}

\def\fdir{f^\dir}
\def\fcas{f^\cas}
\def\fmis{f^\mis}

\def\Tmis{\T^\mis}

\newcommand{\BABARPubNumber}  {05/13}

\newcommand{\SLACPubNumber} {11136}

\def\figurebox#1#2#3{%
    \def\arg{#3}%
    \ifx\arg\empty
    {\hfill\vbox{\hsize#2\hrule\hbox to #2{\vrule\hfill\vbox to #1{\hsize#2\vfill}\vrule}\hrule}\hfill}%
    \else
    {\hfill\epsfbox{#3}\hfill}%
    \fi}

\begin{document}

\preprint{\babar-PUB-02/013}
\preprint{SLAC-PUB-\SLACPubNumber}

\begin{flushleft}
\babar-PUB-\BABARPubNumber\\
SLAC-PUB-\SLACPubNumber\\
\end{flushleft}

\title{\large \bf 
Measurement of 
Time-Dependent \boldmath \CP-Violating Asymmetries and Constraints on $\sin(2\beta+\gamma)$ with
Partial Reconstruction of $\btodstpipm$ Decays

}
  

%
\author{B.~Aubert}
\author{R.~Barate}
\author{D.~Boutigny}
\author{F.~Couderc}
\author{Y.~Karyotakis}
\author{J.~P.~Lees}
\author{V.~Poireau}
\author{V.~Tisserand}
\author{A.~Zghiche}
\affiliation{Laboratoire de Physique des Particules, F-74941 Annecy-le-Vieux, France }
\author{E.~Grauges}
\affiliation{IFAE, Universitat Autonoma de Barcelona, E-08193 Bellaterra, Barcelona, Spain }
% ...
\collaboration{The \babar\ Collaboration}
\noaffiliation

\date{\today}

\begin{abstract}
We present a  
measurement of the time-dependent \CP-violating asymmetries in
decays of neutral $B$ mesons to the final states $\Dstarmp\pi^\pm$,
using approximately 
$232$ million $\BB$ events recorded by the \babar\ experiment
at the \pep2\ $\epem$ storage ring.  Events containing these decays are
selected with a partial reconstruction technique, in which only the
high-momentum $\pi^\pm$ from the $B$ decay and the low-momentum 
$\pi^\mp$ from the $\Dstarmp$ decay are
used. 
We measure the parameters related to $2 \beta + \gamma$ to be 
$a_\dstpi=-0.034 \pm 0.014 \pm 0.009$ and
$c_\dstpi^\ell = -0.019 \pm 0.022 \pm 0.013$.
With some theoretical assumptions, we
interpret our results in terms of the lower limits
$|\sinphi|> 0.62~(0.35)$ at 68\% (90\%) confidence level. 
\end{abstract}
 
 \maketitle
\vskip .3 cm
 
\setcounter{footnote}{0}


\section{INTRODUCTION}
\label{sec:Introduction}

\begin{figure}[hp]
\begin{center}
\begin{tabular}{cc}
        \includegraphics[width=0.24\textwidth]{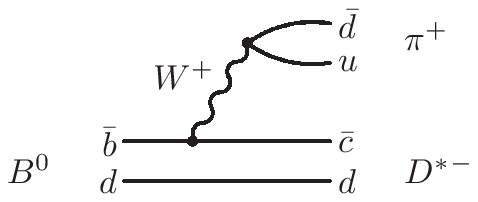}
&
        \includegraphics[width=0.24\textwidth]{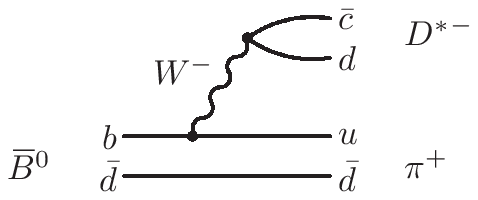}
\end{tabular}
\end{center}
\caption{Feynman diagrams for the Cabibbo-favored decay
$\Bz \rightarrow \Dstarm\pi^+$ (left), corresponding to the
decay amplitude $A_c$, and the Cabibbo-suppressed decay
$\Bzb \rightarrow \Dstarm\pi^+$ (right), whose amplitude is $A_u$.}
\label{fig:dstpi-diagrams}
\end{figure}

The Cabibbo-Kobayashi-Maskawa (CKM)
quark-mixing matrix~\cite{ref:km} provides an 
explanation of \CP violation
and is under 
experimental investigation aimed at constraining its parameters. A
crucial part of this program is the measurement of the angle $\gamma =
\arg{\left(- V^{}_{ud} V_{ub}^\ast/ V^{}_{cd} V_{cb}^\ast\right)}$ of
the unitarity triangle related to the CKM matrix.
The decay modes $B \rightarrow {\Dstar}^{\mp} \pi^{\pm}$ have been
proposed for use in measurements of
$\sin(2\beta+\gamma)$~\cite{ref:book}, where $\beta = \arg{\left(-
V^{}_{cd} V_{cb}^\ast/ V^{}_{td} V_{tb}^\ast\right)}$ is well
measured~\cite{ref:sin2b}.
In the Standard Model the decays 
$\Bz \to \Dstarm \pi^+$ and $\Bzb \to \Dstarm \pi^+$
proceed through the $\overline{b} \rightarrow \overline{c}  u
\overline{d}  $ and
$\btou$ amplitudes $A_c$ and $A_u$. 
Fig.~\ref{fig:dstpi-diagrams} shows the tree diagrams contributing to these
decays. 
The relative weak phase between $A_u$ and $A_c$
in the usual Wolfenstein convention~\cite{ref:wolfen}
is $\gamma$.
When combined with $\Bz \Bzb$ mixing, this yields a weak phase
difference of $2\beta+\gamma$ between the interfering amplitudes.

In $\Upsilon(4{\rm S})\to \BB$ decays, the decay rate distribution for 
$B \to {\Dstar}^\mp\pi^\pm$ is
\begin{eqnarray}
\label{eq:pure-dt-pdf-B}
\P^\pm_\eta(\dt)
&=& {e^{-|\dt|/\tau} \over 4\tau} \times       
\left[ 1 \mp S^\zeta \sin(\Delta m \dt) \right.  \nonumber\\  
& & \left.\mp \eta C \cos(\Delta m \dt) \right],
\end{eqnarray}
where  
$\tau$ is the $\Bz$ lifetime averaged over the two mass eigenstates,
$\Delta m$ is the $\Bz \Bzb$ mixing 
frequency, and $\dt$
is the difference between the time
of the $B\to{\Dstar}^\mp\pi^\pm$ ($\Brec$)
decay and the decay of the other
$B$ ($\Btag$) in the event. The
upper (lower) signs in Eq.~(\ref{eq:pure-dt-pdf-B})
indicate the flavor of the $\Btag$ as a $\Bz$ ($\Bzb$),
while $\eta = +1$ ($-1$) and $\zeta = +$ ($-$) for
the $\Brec$ final state ${\Dstar}^-\pi^+$ (${\Dstar}^+\pi^-$).
The parameters $C$ and $S^\pm$ are given by
\begin{equation}
C \equiv {1 - \rsq \over 1 + \rsq}\, , \ \ \ \ 
S^\pm \equiv {2 \r \over 1 + \rsq}\, \sin(2 \beta + \gamma \pm \deltaPhase).
\label{eq:AandB}
\end{equation}
Here $\deltaPhase$ is the strong phase difference 
between $A_u$  and $A_c$,
and 
$\r = | A_u / A_c |$.
Since $A_u$ is doubly CKM-suppressed with respect
to $A_c$, one expects $\r\approx \left| \frac{V_{ub}^{} V_{cd}^\ast} 
{V_{cb}^\ast V_{ud}^{}} \right| = 0.02$. 

We report a study of the \CP-violating asymmetry in $\btodstpipm$
decays using the technique of partial reconstruction, which allows us
to achieve a high efficiency for the  selection of signal events.
We use approximately twice the integrated luminosity of our
previous analysis of this process~\cite{ref:run1-2}, 
and employ
an improved method to eliminate a measurement bias,
as described in Sec.~\ref{sec:signalpdf}.
Many of the tools and procedures used in this analysis were validated
in a previous analysis dedicated to the measurement of the 
$\Bz$ lifetime \cite{ref:dstpi-lifetime}.

In this analysis, terms of order $r^{*2}$, to which we currently have no
sensitivity, have been neglected. The interpretation of the measured asymmetries 
in terms of $\sin(2\beta + \gamma)$ requires an assumption regarding the value 
of $\r$, discussed in  Sec.~\ref{sec:Physics}.

\section{THE \babar\ DETECTOR AND DATASET}
\label{sec:babar}

The data used in this analysis were recorded with the \babar\
detector at the \pep2\ asymmetric-energy storage rings, and consist of 211~fb$^{-1}$
collected on the $\Upsilon(4{\rm S})$ resonance (on-resonance
sample), and 21~fb$^{-1}$ collected at an $\epem$ center-of-mass (CM) 
energy approximately 40~\mev below the resonance peak 
(off-resonance sample). Samples of Monte Carlo (\mc)~\cite{ref:geant4} events
with an equivalent luminosity approximately four times larger than the data 
sample were
analyzed using the same reconstruction and analysis procedure.

The \babar\ detector is described in detail in Ref.~\cite{ref:babar}.
We provide a brief description of the main components and their use in
this analysis.  Charged-particle trajectories are measured by a
combination of a five-layer silicon vertex tracker (SVT) and a
40-layer drift chamber (DCH) in a 1.5-T solenoidal magnetic field.
Tracks with low transverse momentum can be reconstructed in the SVT
alone, thus extending the charged-particle detection down to
transverse momenta of about  50~\mevc. We use a ring-imaging Cherenkov 
detector (DIRC) for charged-particle identification and  augment it with  
energy-loss measurements from the
SVT and DCH. Photons and electrons are
detected in a CsI(Tl) electromagnetic calorimeter (EMC), with 
photon-energy resolution $\sigma_E / E = 0.023 (E/\gev)^{-1/4} \oplus
0.014$. The instrumented flux return (IFR) is equipped with
resistive plate chambers to identify muons.

\section{ANALYSIS METHOD}
\label{sec:Analysis}

\subsection{Partial Reconstruction of \boldmath $\btodstpipm$}
\label{sec:partial}

In the partial reconstruction of a $\btodstpipm$ candidate ($\Brec$), only the hard
(high-momentum) pion track $\pi_h$  from the $B$ decay and the
soft (low-momentum) pion track $\pi_s$ from the decay 
$D^{*-}\rightarrow \Dzb \pi_s^-$ are used.
The cosine of the angle between the momenta
of the $B$ and the hard pion in the CM frame is then computed:
\begin{equation}
\cos\theta_{Bh} =
        {M_{{\Dstar}^-}^2 - M_{\Bz}^2 - M_{\pi}^2 + E_{\rm CM} E_h
                        \over
         2 p_B |\vec p_{h}|
        },
\label{eq:cosTheta}
\end{equation}
where $M_x$ is the nominal mass of particle~$x$~\cite{ref:pdg2004},
$E_h$ and $\vec p_h$
are the measured CM energy and momentum of the hard pion, $E_{\rm CM}$ is
the total CM energy of the incoming \epem\ beams, and
$p_B = \sqrt{E_{\rm CM}^2/4 - M_{\Bz}^2}$.
Events are required to be in the
physical region  $|\cos\theta_{Bh}|<1$.
Given $\cos\theta_{Bh}$ and the measured momenta of the $\pi_h$ and $\pi_s$,
the $B$ four-momentum can be calculated up to
an unknown azimuthal angle $\phi$ around ${\vec p}_{h}$. For every
value of $\phi$, the expected $D$ four-momentum $
p_D(\phi)$ is determined from four-momentum conservation, and the
corresponding
$\phi$-dependent invariant mass
$m(\phi) \equiv \sqrt{|p_D(\phi)|^2}$
is calculated.
We define the missing mass
$\mmiss \equiv {1 \over 2}\left[m_{\rm max} + m_{\rm min}\right]$,
where $m_{\rm max}$ and $m_{\rm min}$ are the maximum and minimum
values of $m(\phi)$. In signal events, $\mmiss$ peaks at
the nominal $\Dz$ mass $M_{\Dz}$, with a gaussian width of about 3~\mevcc (Fig.~\ref{fig:data_mmiss}).
The $\mmiss$ distribution
for combinatoric background events is significantly broader, making the
missing mass the primary variable for distinguishing signal from
background.
The  discrimination between signal and background 
provided by the $\mmiss$ distribution
is independent of the choice of the value of $\phi$.
With the arbitrary choice
$\phi=0$, we use four-momentum conservation to
calculate the CM $D$ and $B$ momentum vectors, which are used as described below.

\subsection{Backgrounds}
\label{sec:bgd}

In addition to $\btodstpipm$  events, the selected event sample  
contains the following kinds of events:
\begin{itemize}
\item $\btodstrhopm$.

\item Peaking $\BB$ background, defined as decays other than $\btodstrhopm$, 
in which the $\pi_h$ and $\pi_s$ originate from
  the same $B$ meson, with the $\pi_s$ originating from a
  charged $\Dstar$ decay. The $\mmiss$ distribution of these events
  peaks broadly under the signal peak.

\item Combinatoric \BB background, defined as all remaining $\BB$ background 
events.

\item  Continuum $\epem \rightarrow \qqbar$, 
where $q$
represents a $u$, $d$, $s$, or $c$ quark.
\end{itemize}

%
%
\subsection{Event Selection}
\label{sec:cuts}
 
To suppress the continuum background, we select events in which the
ratio of the 2nd to the 0th Fox-Wolfram moment~\cite{ref:R2}, computed
using all charged particles and EMC clusters not matched to tracks, is
smaller than 0.40.
Hard-pion candidates are required to be reconstructed with at least
twelve DCH hits. Kaons and leptons are rejected from the $\pi_h$
candidate lists based on
information from the IFR and DIRC, energy loss in the SVT and DCH, or
the ratio of the
candidate's EMC energy deposition to its momentum ($E/p)$.

We define the $\Dstar$ helicity angle $\theta_{\Dstar}$ to be the
angle between the flight directions of the $D$ and the $B$ in the
$\Dstar$ rest frame.
Taking advantage of the longitudinal polarization in signal events,
we suppress background by requiring $| \cos \theta_{\Dstar} |$ to be
larger than 0.4.

All candidates are required to satisfy $\mmiss >
1.81$~\gevcc. 
Multiple candidates are found in 5\% of the events. In these instances,
only the candidate with the $\mmiss$ value closest to $M_{\Dz}$ is used.

\subsection{Fisher Discriminant}
\label{sec:fisher}

To further discriminate against continuum events,
we combine fifteen event-shape 
variables into a Fisher discriminant~\cite{ref:fisher} $F$.
Discrimination originates from the fact that $\qqbar$ events tend to
be jet-like, whereas $\BB$ events have a more spherical energy
distribution. Rather than applying requirements to the variable $F$,
we maximize the sensitivity  by using it in the fits described below.
The fifteen variables are calculated using two sets of particles.
Set~1 includes all tracks and EMC clusters, excluding the hard and
soft pion candidates; Set~2 is composed of Set~1, excluding all tracks
and clusters with CM momentum within 1.25~radian of the CM momentum of the
$D$. The variables, all calculated in the CM
frame, are
1) the scalar sum of the momenta of all Set~1 tracks and EMC
clusters in nine
$20^{\circ}$ angular bins centered about the hard pion direction;
2) the value of the sphericity, computed with Set~1;
3) the angle with respect to the hard pion of the sphericity axis, computed
with Set~2;
4) the direction of the particle of highest energy in Set~2
with respect to the hard pion;
5) the absolute value of the vector sum of the momenta of
all the particles in Set~2;
6) the momentum $|\vec p_h|$ of the hard pion and its polar angle.

\subsection{Decay Time Measurement and Flavor Tagging}
\label{sec:deltat}

To perform this analysis, $\dt$ and the flavor of the $\Btag$ must be
determined. 
We tag the flavor of the $\Btag$ using lepton or kaon
candidates.
The lepton CM momentum is required to be greater than 1.1~\gevc 
to suppress 
leptons that originate from charm decays.
If several flavor-tagging tracks are present in either the lepton or kaon
tagging category,
the only track of that category used for tagging is the one with the
largest value of $\theta_T$, the CM angle between the track
momentum and the momentum of the ``missing'' (unreconstructed) $D$.  
The tagging track must satisfy
$\cos \theta_T<C_T $, 
where $C_T=0.75 $ ($C_T=0.50 $) for leptons (kaons),  to minimize
the impact of tracks originating from the decay of the missing 
$D$. If both a lepton and a
kaon satisfy this requirement, the event is tagged with the lepton.

We measure $\dt$ using $\dt = (\zrec - \ztag) /
(\gamma\beta c)$, where $\zrec$ ($\ztag$) is the decay position of the
$\Brec$ ($\Btag$) along the beam axis ($z$) in the laboratory frame,
and the $e^+e^-$ boost parameter $\gamma\beta$ is
calculated from the measured beam energies. 
To find $\zrec$, we use the $\pi_h$ track parameters and errors,
and the measured beam-spot position and size in the plane perpendicular to the
beams (the $x - y$ plane). We find the position of the point in space
for which the sum of the $\chi^2$ contributions from the $\pi_h$ track
and the beam spot is a minimum. The $z$ coordinate of this point
determines $\zrec$.
The beam spot has an r.m.s. size of approximately 120~$\mu$m in the horizontal
dimension ($x$), 5~$\mu$m in the vertical dimension ($y$), and
8.5~mm along the beams ($z$). The average $B$ flight in the $x-y$ plane is 
30~$\mu$m. 
To account for the $B$ flight in the
beam-spot-constrained vertex fit, 30~$\mu$m are added to the effective 
$x$ and $y$ sizes for the purpose of conducting this fit.

In lepton-tagged events, the same procedure,
with the $\pi_h$ track replaced by the tagging lepton, is used to
determine $\ztag$.

In kaon-tagged events, we obtain $\ztag$ from a beam-spot-constrained
vertex fit of all tracks in the event, excluding $\pi_h$, $\pi_s$ and all tracks
within 1~radian of the $D$ momentum in the CM frame.
If the contribution of any track to the $\chi^2$ of the vertex 
is more than 6, the track is removed and the fit is repeated until
no track fails the $\chi^2 < 6$ requirement.

The $\dt$ error $\dtErr$ is calculated from the results
of the $\zrec$ and $\ztag$ vertex fits. We require $|\dt| < 15 \ps$ and 
 $\dtErr < 2 \ps$.

\subsection{Probability Density Function}
\label{sec:pdf}

The probability density function (PDF) depends on the variables
$\mmiss$, $\dt$, $\dtErr$, $F$, $\stag$, and $\smix$,
where $\stag = 1$ ($-1 $) when the $\Btag$ is identified as a $\Bz$ ($\Bzb$), 
and $\smix = 1$ ($-1 $) for ``unmixed'' (``mixed'') events.
An event is labeled unmixed if the $\pi_h$ is a 
$\pi^- (\pi^+)$ and the $\Btag$ is a $\Bz (\Bzb)$, and mixed
otherwise.

The PDF for on-resonance data is a sum over the PDFs of
the different event types:
\begin{equation}
\P = \sum_{i}f_{i} \, \P_i,
\label{eq:pdf-sum}
\end{equation}
where the index $i = \{\dstpi, \dstrho, \peak, \comb, \cont\}$
indicates one of the event types described above, $f_i$ is the
relative fraction of events of type $i$ in the data sample, and $\P_i$
is the PDF for these events.
The PDF for off-resonance data is $\P_\cont$.
The parameter values for $\P_i$ are different for each event type,
unless indicated otherwise.  Each $\P_i$ is a product,
\beq 
\P_i = \M_i(\mmiss)\, \F_i(F)\, \T'_i(\dt, \dtErr, \stag, \smix),
\label{eq:pdf-prod}
\eeq
where the factors in Eq.~(\ref{eq:pdf-prod}) are described below.

\subsubsection{ $\mmiss$ and $F$ PDFs }

The $\mmiss$ PDF for each event type $i$ is the sum of a bifurcated
Gaussian plus an ARGUS function~\cite{ref:argus}:
\begin{equation}
\M_i(\mmiss) = f^{\BG}_i\, \BG_i(\mmiss) + (1-f^{\BG}_i) \A_i(\mmiss), 
\label{eq:mmiss-pdf}
\end{equation}
where $f^{\BG}_i$ is the fractional area of the bifurcated Gaussian function. 
The functions $\BG_i$ and $\A_i$ are
\begin{eqnarray}
&& \kern-0.6cm \BG_i(m) \propto \left\{ \begin{matrix}
        \exp\left[-(m - M_i)^2 / 2\sigma_{Li}^2\right] & , & 
                                        m \le M_i \\[2mm]
     \exp\left[-(m - M_i)^2 / 2\sigma_{Ri}^2\right] & , & 
                                        m > M_i 
                            \end{matrix}\right. , 
\label{eq:bifur}
\\[2mm]
&& \kern-0.6cm \A(m) 
        \propto m \sqrt{1-\left({m /M^A_i}\right)^2}\;
\times \nonumber \\
&&   \exp\left[\epsilon_i \left(1-\left({m / M^A_i}\right)^2\right)\right] 
 \theta( M^A_i-m),
\label{eq:argus}
\end{eqnarray}
where $M_i$ is the peak of the bifurcated Gaussian, $\sigma_{Li}$ and
$\sigma_{Ri}$ are its left and right widths, $\epsilon_i$ is
the ARGUS exponent, $M^A_i$ is its end point, and 
$\theta$ is the step function. 
The proportionality
constants are such that each of these functions is normalized to unit
area within the $\mmiss$ range. 
The $\mmiss$ PDF of each event type has different parameter values.


The Fisher discriminant PDF $\F_i$ for each event type is
parameterized as the sum of two Gaussians. 
The parameter values of $\F_\dstpi$, $\F_\dstrho$, $\F_\peak$, and
$\F_\comb$ are identical.

\subsubsection{ Signal  $\dt$ PDFs}
\label{sec:signalpdf}

The $\dt$ PDF $\T'_\dstpi(\dt, \dtErr, \stag, \smix)$ for signal events
corresponds to Eq.~\ref{eq:pure-dt-pdf-B} with $O(\rsq)$ 
terms neglected, modified to  account
for several experimental effects, described below.

The first effect has to do with the origin of the tagging track.
In some of the events, the tagging track
originates from the decay of the missing $D$.
These events are labeled ``missing-$D$ tags'' and do not 
provide any information regarding the flavor of the $\Btag$.
In lepton-tagged events, we further distinguish between ``direct'' tags,
in which the tagging lepton originates directly from the decay of the
$\Btag$, and ``cascade'' tags, where the tagging lepton is a daughter
of a charmed particle produced in the $\Btag$ decay. Due to the different
physical origin of the tagging track in cascade and
direct tags, these two event categories
have different mistag probabilities, defined as the probability to deduce
the wrong $B$ flavor from the charge of the tagging track.  
In addition,  the measured value
of $\ztag$ in cascade-lepton tags is systematically larger than the
true value, due to the finite lifetime of the charmed particle and the
boosted CM frame.  This creates a correlation between the tag and
vertex measurements that we address by considering
cascade-lepton tags separately in the PDF.  
  In our previous analysis~\cite{ref:run1-2} we corrected for the bias 
of the 
$S^{\pm}$
parameters caused by this
  effect and included a systematic error due to its uncertainty.
In kaon tags, $\ztag$ is determined using all available $\Btag$
tracks, so the effect of the tagging track on the $\ztag$
measurement is small. 
Therefore, the overall bias induced by cascade-kaon tags is small,
and there is no need to distinguish them in the PDF.

The second experimental effect is the finite detector resolution in the
measurement of $\dt$. We address this by convoluting the distribution
of the true decay time difference $\dttrue$ with a detector resolution
function. Putting these two effects together, the $\dt$ PDF of signal
events is
\bea
&& \T'_\dstpi (\dt, \dtErr, \stag, \smix) = (1 + \stag \, \Delta\epsilon_\dstpi) 
	\sum_j f^j_\dstpi \times \nonumber  \\
        && \int d\dttrue\, \T_\dstpi^j(\dttrue, \stag, \smix) \,
        \R_\dstpi^j(\dt - \dttrue, \dtErr),
\label{eq:CP-pdf-sig}
\eea
where
$ \Delta\epsilon_\dstpi$ is half the relative difference between 
the detection efficiencies of positive and negative leptons or kaons,
the index $j = \{\dir, \ \cas, \ \mis\}$ indicates direct, cascade, and
missing-$D$ tags,
and $f^j_\dstpi$ is the fraction of signal events of tag-type $j$ in
the sample.  We set $\fdir_\dstpi = 1 - \fcas_\dstpi -
\fmis_\dstpi$ for lepton tags, with the value 
$\fcas_\dstpi=0.12 \pm 0.02$ obtained from the MC simulation. 
For kaon tags $\fdir_\dstpi = 0$.
The function $\T_\dstpi^j(\dttrue, \stag, \smix)$ is the $\dttrue$ distribution
of tag-type $j$ events,
and $\R_\dstpi^j(\dt - \dttrue, \dtErr)$ 
is their resolution function, which parameterizes 
both the finite detector resolution and systematic 
offsets in the measurement of $\dz$, such as those due to the 
origin of the tagging particle. The parameterization of the resolution
function is described in Sec.~\ref{sec:res}. 


The functional form of the direct and cascade tag $\dttrue$ PDFs is
\bea
\T^j_\dstpi(\dttrue, \stag, \smix)
	&=&  {e^{-|\dttrue|/\tau_\dstpi}\over 4 \tau_\dstpi} \times    \nonumber \\
	&& \kern-0.6cm
		\biggl\{1-\stag\,\Delta\omega^j_\dstpi
	\nonumber\\
        &&  \kern-0.6cm + \smix\,(1-2\omega^j_\dstpi)\, \cos(\Delta m_\dstpi
          \dttrue)      
	\nonumber\\   
        &&  \kern-0.6cm  - {\cal S}^j_\dstpi 
                \,\sin(\Delta m_\dstpi \dttrue) 
         \biggr\},
\label{eq:CP-pdf-dir-cas}
\eea
where 
$j = \{\dir, \ \cas\}$, the mistag rate $\omega^j_\dstpi$ is the probability to
misidentify the flavor of the $\Btag$ averaged over $\Bz$ and $\Bzb$,
and $\Delta \omega^j_\dstpi$ is the 
$\Bz$ mistag rate minus the $\Bzb$ mistag rate.
The factor ${\cal S}^j_\dstpi$ describes 
the effect of interference 
between $\btou$ and $\btoc$ amplitudes in both 
the $\Brec$ and the 
$\Btag$ decays:
\bea
{\cal S}^j_\dstpi= && (1-2\omega^j_\dstpi) \, (\stag a_\dstpi + \smix c_\dstpi) \nonumber \\
        && + \stag \smix b_\dstpi (1-\stag \Delta\omega^j_\dstpi) ,
\label{eq:cp-pdf-sin}
\eea
where $a_\dstpi$, $b_\dstpi$, and $c_\dstpi$ are related to the physical
parameters through
\bea
a_\dstpi&\equiv& 2 \r\sin(2\beta+\gamma)\cos\deltaPhase , \nonumber\\
b_\dstpi&\equiv& 2 \rp\sin(2\beta+\gamma)\cos\deltaPhaseP , \nonumber\\
c_\dstpi&\equiv& 2\cos(2\beta+\gamma)(\r\sin\deltaPhase -\rp\sin\deltaPhaseP), 
\label{eq:abc}
\eea
and $\rp$ ($\deltaPhaseP$) is the effective magnitude of the ratio 
(effective strong phase
difference) between the $b\rightarrow u \overline c
d$ and $b\rightarrow c \overline u d$ amplitudes in the $\Btag$ 
decay. 
This parameterization is good to first order in $\r$ and $\rp$.
In the following we will refer to the parameters 
$a_\dstpi$, $b_\dstpi$, $c_\dstpi$ and related parameters for the 
background PDF 
as the weak phase parameters. Only $a_\dstpi$ and $b_\dstpi$ are
related to \CP violation, while $c_\dstpi$ can be non-zero even in the
absence of \CP violation when $2\beta+\gamma=0$.
The inclusion of $\rp$ and $\deltaPhaseP$ in the 
formalism accounts for cases where the 
$\Btag$ undergoes a $b\rightarrow u \bar c d$ decay, and the kaon 
produced in the subsequent charm decay is used for tagging~\cite{ref:abc}.
We expect $\rp \sim  0.02$.
In lepton-tagged events $\rp = 0$ (and hence $b_\dstpi=0$) because 
most of the tagging leptons come from $B$ semileptonic decays to which 
no suppressed amplitude with a different weak phase can contribute.

The $\dttrue$ PDF for missing-$D$ tags is
\begin{eqnarray}
\Tmis_\dstpi(\dttrue, \stag, \smix) &=& 
	{e^{-|\dttrue|/\tau^\mis_\dstpi}\over 8 \tau^\mis_\dstpi} 
	\biggl\{1 + \smix \left( 1 - 2\rho_\dstpi\right) 
 \nonumber \\ 
		&-&2 \stag \smix b_\dstpi \,\sin(\Delta m_\dstpi \dttrue) 
\biggr\},
\label{eq:CP-pdf-mis}
\end{eqnarray}
where $\rho_\dstpi$ is the probability that the charge of the tagging track
is such that it results in a mixed flavor measurement. 
In this analysis, we have neglected the term proportional to 
$\sin(\Delta m_\dstpi \dttrue)$ of Eq.~\ref{eq:CP-pdf-mis}.
The systematic error on $b_\dstpi $ due to this 
approximation is negligible due to the small value of $\fmis_\dstpi$  
reported below.

\subsubsection{Background  $\dt$ PDFs}
\label{sec:bgd-pdfs}

The $\dt$ PDF of $\btodstrhopm$ has the same functional form and
parameter values as the signal PDF, except that the weak phase parameters
$a_\dstrho$, $b_\dstrho$, and $c_\dstrho$ are set to 0 and are later
varied to evaluate systematic uncertainties.
The validity of the use of the same parameters 
for $\T'_\dstrho$ and 
$\T'_\dstpi$ is established using simulated events, and
stems from the fact that the $\pi_h$ momentum spectrum in the 
$\btodstrhopm$  events that pass our selection criteria is almost 
identical
to the signal spectrum.

The $\dt$ PDF of the peaking background accounts separately for 
charged and neutral $B$ decays:
\bea
&& \kern-0.6cm \T'_\peak (\dt, \dtErr, \stag, \smix) = (1 + \stag \, \Delta\epsilon_\peak) 
	\left\{\T{^0}'_\peak  \right.	\nonumber\\
	&& \kern-0.6cm 
        + \int d\dttrue\, \T_\peak^+(\dttrue, \stag, \smix) \, \times  
\nonumber\\
&& \left.       \R_\peak^+(\dt - \dttrue, \dtErr) \right\},	
\label{eq:pdf-peak}
\eea
where $\T{^0}'_\peak$ has the functional form of
Eq.~(\ref{eq:CP-pdf-sig}) and the subsequent expressions,
Eqs.~(\ref{eq:CP-pdf-mis}-\ref{eq:abc}), but with all $\dstpi$-subscripted
parameters replaced with their $\peak$-subscripted counterparts. 
The integral in Eq.~(\ref{eq:pdf-peak}) accounts for 
the contribution of charged $B$ decays to the peaking background,
with 
\beq
\T_\peak^+(\dttrue, \stag) =  
	{e^{-|\dttrue|/\tau^+_\peak}\over 4 \tau^+_\peak}        
        \left(1-\stag\,\Delta\omega^+_\peak \right),
\label{eq:pdf-peak-charged}
\eeq
and $\R_\peak^+(\dt - \dttrue, \dtErr)$ being the three-Gaussian resolution 
function for these events described below.
%

The Combinatoric $\BB$ background PDF $\T'_\comb$ is similar to 
the signal PDF, with one substantial difference.
Instead of parameterizing $\T'_\comb$ with the four parameters
$\fdir_\comb$, $\omega^\dir_\comb$, $\Delta\omega^\dir_\comb$, $\rho_\comb$,
we use the set of three parameters
\begin{eqnarray}
  \omega'_\comb &=& 
     \omega_\comb^\dir\,(1-\fdir_\comb)+\frac{\fdir_\comb}{2}, \nonumber\\
  \Delta\omega'_\comb &= &  
      \Delta\omega_\comb\,(1-\fdir_\comb),  \nonumber\\
  {\Omega_\comb} & = & \fdir_\comb (1-2\,\rho_\comb). 
\end{eqnarray}
With these parameters and $\fcas_\comb = 0$, the 
combinatoric $\BB$ background $\dt$ PDF becomes 
\bea
&& \kern-0.7cm \T'_\comb (\dt, \dtErr, \stag, \smix) =   (1 + \stag \, \Delta\epsilon_\comb) \times \nonumber \\
        && \kern-0.7cm \int d\dttrue\, \T_\comb(\dttrue, \stag, \smix) \,
        \R_\comb(\dt -\dttrue,\dtErr) \,,	
\label{eq:pdf-comb}
\eea
where $\R_\comb(\dt - \dttrue, \dtErr)$ is the 3-Gaussian resolution function
and
\bea
&&\kern-0.6cm  \T_\comb(\dttrue,\stag, \smix)=  {e^{-|\dttrue|/\tau_\comb}\over 4 \tau_\comb} 
	\biggl\{   1-\stag\,\Delta\omega'_\comb  \nonumber\\        
	 &&\kern-0.6cm  
	+ \smix \Omega_\comb 	+ \smix\,(1-2\omega'_\comb)\, \cos(\Delta m_\comb
          \dttrue)         
	\nonumber\\
         &&\kern-0.6cm 
	- {\cal S}_\comb 
                \,\sin(\Delta m_\comb \dttrue) 
         \biggr\},
\label{eq:CP-pdf-comb}
\eea
with
\bea
{\cal S}_\comb &=&  (1-2\omega'_\comb) \, (\stag a_\comb + \smix c_\comb)  \nonumber \\ 
        &+& \stag \smix b_\comb (1-\stag \Delta\omega'_\comb) .
\label{eq:cp-pdf-sin-comb}
\eea

As in the case of $\T_\dstrho$, the weak phase parameters of the peaking 
and combinatoric background ($a_\peak$, $b_\peak$, $c_\peak$ 
and $a_\comb$, $b_\comb$, $c_\comb$) are set to 0
and are later varied to evaluate systematic uncertainties.
Parameters labeled with superscripts ``$\peak$'' or ``$\comb$''
are empirical and thus do not necessarily correspond to 
physical parameters. In general, their values may be different from those of 
the $\dstpi$-labeled parameters.

%
The PDF $\T_\cont$ for the continuum background is
the sum of two components, one with a finite lifetime
and one with zero lifetime:
\bea
\T'_\cont (\dt, \dtErr, \stag) &=& (1+\stag \, \Delta\epsilon_\cont)
        \int d\dttrue\, \T_\cont(\dttrue, \stag, \smix) \nonumber \\
        &\times& \R_\cont(\dt - \dttrue, \dtErr) ,	
\label{eq:pdf-qq}
\eea
with 
\bea
\T_\cont(\dttrue, \stag) &=& (1-f^\delta_\cont)
	{e^{-|\dttrue|/\tau_\cont}\over 4 \tau_\cont}        
        \left(1-\stag\,\Delta\omega_\cont \right)\nonumber \\
	 &+& f^\delta_\cont\, \delta(\dttrue),
\label{eq:pdf-qq-2}
\eea
where $f^\delta_\cont$ is the fraction of zero-lifetime events.

\subsubsection{Resolution Function Parameterization}
\label{sec:res}

The resolution function for events of type $i$ and optional secondary type
$j$ ($j = \{\dir, \ \cas, \ \mis\}$ for lepton-tagged signal events and $j = \{+, \ 0\}$ 
for the peaking and combinatoric $\BB$ background types)
is parameterized as the sum of
three Gaussians:
\bea
\R\ij(t_r, \dtErr) &=& 
        f^n\ij\, \G^n\ij(t_r, \dtErr) \nonumber \\
	&+& (1 - f^n\ij - f^o\ij)\, \G^w\ij(t_r, \dtErr) \nonumber \\ 
        &+& f^o\ij\, \G^o\ij(t_r, \dtErr),
\label{eq:res}
\eea
where $t_r = \dt - \dttrue$ is the residual of the $\dt$
measurement, and $\G^n\ij$, 
$\G^w\ij$, and
$\G^o\ij$ are the ``narrow'', ``wide'', and ``outlier'' Gaussians. The
narrow and wide Gaussians have the form
%
\begin{eqnarray}
\G^k\ij(t_r, \dtErr) &\equiv& 
        {1 \over \sqrt{2\pi} \, s^k\ij\, \dtErr}
\times \nonumber \\
&&  \exp\left(-\,{\left(t_r - b^k\ij\dtErr\right)^2  
        \over 2 (s^k\ij\, \dtErr)^2}\right), 
\label{eq:Gaussians}
\end{eqnarray}
where the index $k$ takes the values $k=n,w$ for the narrow and wide
Gaussians, and $b^k\ij$ and $s^k\ij$ are parameters determined by
fits, as described in Sec.~\ref{sec:proc}.
The outlier Gaussian has the form 
\begin{equation}
\G^o\ij(t_r, \dtErr) \equiv 
        {1 \over \sqrt{2\pi} \, s^o\ij  }
  \exp\left(-\,{\left(t_r - b^o\ij \right)^2  
        \over 2 (s^o\ij)^2}\right),
\label{eq:GaussiansOutlier}
\end{equation}
where in all nominal fits the values of $b^o\ij$ and $s^o\ij$ are
fixed to 0~ps and 8~\ps, respectively, and are later varied to
evaluate systematic errors.

\subsection{Analysis Procedure}
\label{sec:proc}

The analysis is carried out with a series of unbinned
maximum-likelihood fits, performed simultaneously on the on- and
off-resonance data samples and independently for the lepton-tagged and
kaon-tagged events.
The analysis proceeds in four steps:
\begin{enumerate}
\item In the first step, we determine the parameters
$f_\dstrho+f_\dstpi$, $ f_\peak$, and 
$f_\comb$ of Eq.~(\ref{eq:pdf-sum}).
In order to reduce the reliance on the simulation, we also obtain in
the same fit the parameters
$f^{\BG}_\cont$ of Eq.~(\ref{eq:mmiss-pdf}), 
$\epsilon_\cont$ of Eq.~(\ref{eq:argus}), $\sigma_{L}$ for the signal 
$\mmiss$ PDF (Eq.~(\ref{eq:bifur})),
and all the parameters of the Fisher discriminant PDFs.
This is done by fitting the data with 
the PDF
\beq
\P_i = \M_i(\mmiss) \, \F_i(F),
\label{eq:pdf-prod-kin}
\eeq
instead of Eq.~(\ref{eq:pdf-prod}); i.e. by ignoring the time dependence.  
The fraction $f_\cont$ of continuum events is determined
from the off-resonance sample and its 
integrated luminosity relative to the on-resonance sample.
All other parameters of the $\M_i$ PDFs and the value 
of $f_\dstpi/(f_\dstpi + f_\dstrho)= 0.87 \pm 0.03$ are obtained
from the \mc\ simulation.

\item In the second step, we repeat the fit of the first step for data
events with $\cos \theta_T \ge C_T$, to obtain the fraction of signal
events in that sample.  Given this fraction and the relative
efficiencies for direct, cascade, and missing-$D$ signal events to
satisfy the $\cos \theta_T<C_T$ requirement, we calculate
$\fmis_\dstpi = 0.011 \pm 0.001 $ for lepton-tagged events 
and $\fmis_\dstpi = 0.055 \pm 0.001 $ for kaon-tagged events.  We also calculate
the value of $\rho_\dstpi$ from the fractions of mixed and unmixed signal
events in the $\cos \theta_T \ge C_T$ sample relative to the $\cos
\theta_T < C_T$ sample.

\item In the third step, we fit the data events in the sideband $1.81 <
\mmiss < 1.84$~\gevcc with the 3-dimensional PDFs of
Eq.~(\ref{eq:pdf-prod}).  The parameters of $\M_i(\mmiss)$ and $\F_i(F)$,
and the fractions $f_i$ are fixed to the values obtained in the
first step.  
From this fit we obtain the parameters of $\T'_{\comb}$, as well as
those of $\T'_{\cont}$.

\item In the fourth step, we fix all the parameter
values obtained in
the previous steps and fit the events in the signal region
$\mmiss > 1.845$~\gevcc, determining the parameters
of $\T'_\dstpi$ and $\T'_{\cont}$.
Simulation studies show that the parameters of $\T'_{\comb}$ are
independent of $\mmiss$, enabling us to obtain them in the sideband fit
(step~3) and then use them in the signal-region fit.
The same is not true of the $\T'_{\cont}$ parameters; hence they
are free parameters in the signal-region fit of the last step.
The parameters of $\T'_{\peak}$ are obtained from the MC simulation.
\end{enumerate}


\section{RESULTS}

\begin{figure}[!htbp]
\begin{center}
   \includegraphics[width=0.48\textwidth]{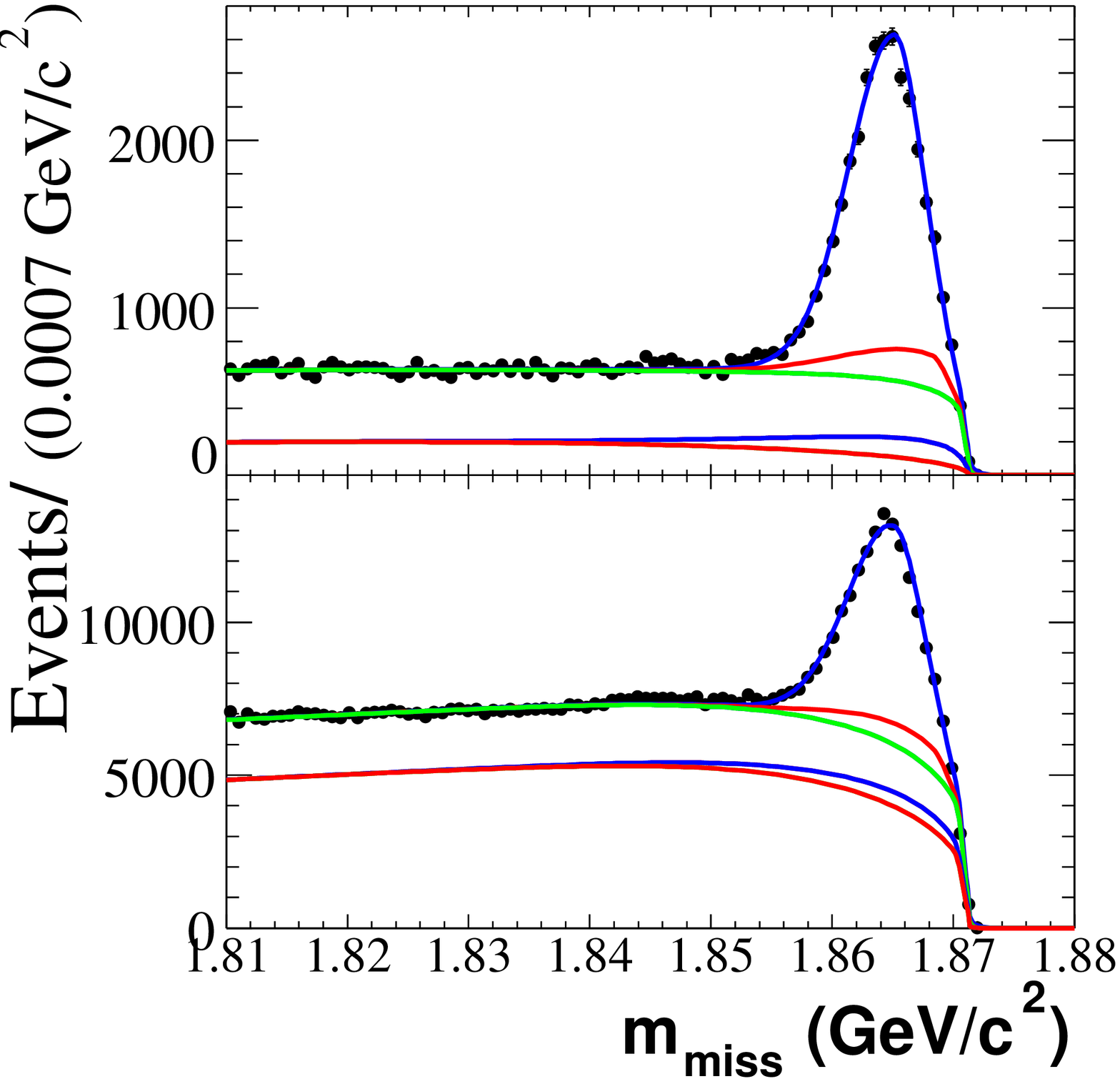}
\end{center}
\caption{The $\mmiss$ 
distributions for on-resonance lepton-tagged
(top) and kaon-tagged (bottom) data. 
The curves show, from bottom
to top, the cumulative contributions of the continuum, peaking \BB, 
combinatoric \BB, $\btodstrhopm$,
and $\btodstpipm$ PDF components.
}
\label{fig:data_mmiss}
\end{figure}
\begin{figure}[!htbp]
\begin{center}
   \includegraphics[width=0.48\textwidth]{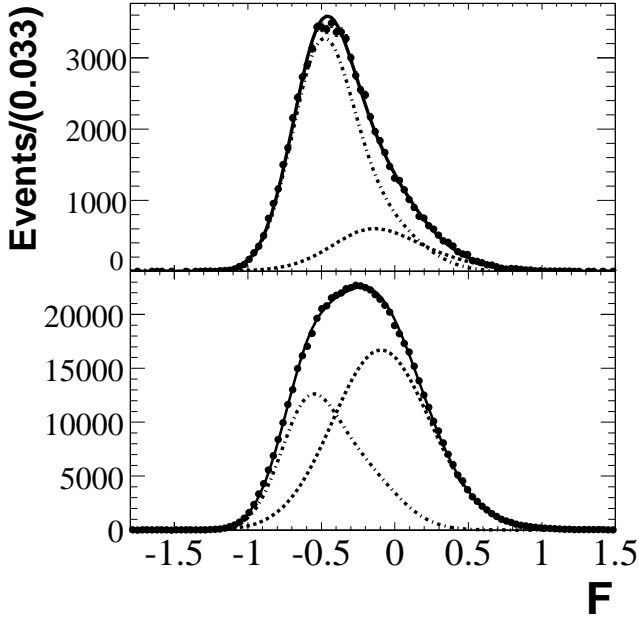}
\end{center}
\caption{The $F$ 
distributions for on-resonance lepton-tagged
(top) and kaon-tagged (bottom) data. 
The contributions of the $\BB$ (dashed-dotted line) and the continuum (dashed
line) PDF components
are overlaid,
peaking at approximately $-0.6$ and $-0.1$, respectively.
The total PDF is also overlaid. 
}
\label{fig:data_fisher}
\end{figure}

The fit of step~1 finds $18710 \pm 270$ signal $\btodstpipm$ events in
the lepton-tag category and $70580 \pm 660$ in the kaon-tag category.  The
$\mmiss$ and $F$ distributions for data are shown in
Figs.~\ref{fig:data_mmiss} and~\ref{fig:data_fisher}, with the PDFs
overlaid.


\begin{figure}[!htb]
\begin{center}
\includegraphics[width=0.48\textwidth]{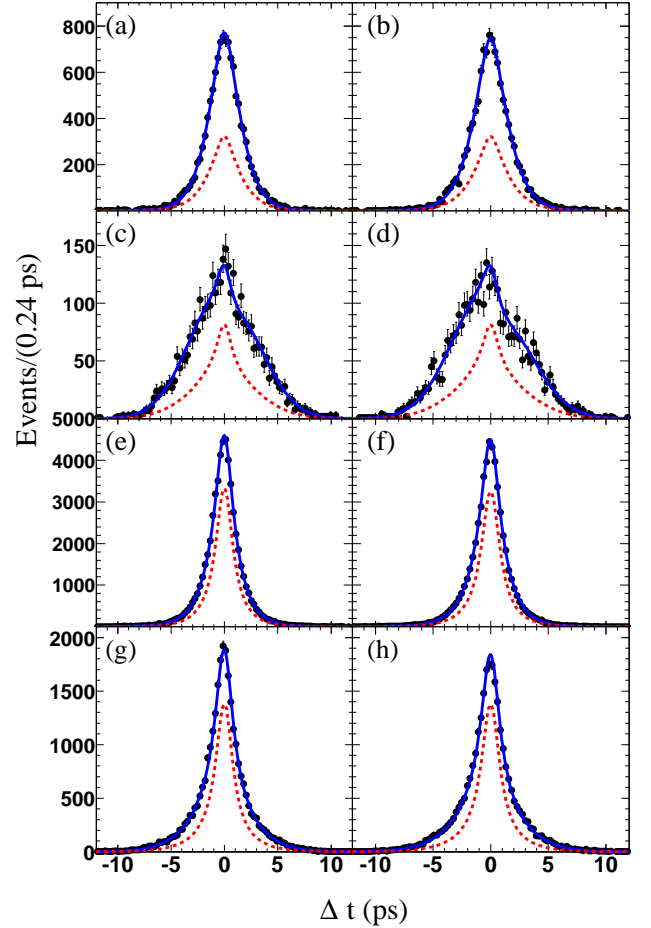}
\end{center}
\caption{$\dt$ distributions for the lepton-tagged (a-d) 
and kaon-tagged (e-h) events 
 separated according to the tagged flavor of $\Btag$ and 
whether they were found to be mixed or unmixed: 
a,e) $\Bz$ unmixed, 
%
b,f) $\Bzb$ unmixed, 
%
c,g) $\Bz$ mixed, 
%
d,h) $\Bzb$ mixed. 
The solid curves show the PDF, calculated with the parameters obtained by the fit.
The PDF for the total background is shown by the dashed curves. 
}
\label{fig:data_sr_Run1-4}
\end{figure}
\begin{figure}
\begin{center}
\begin{tabular}{cc}
   \includegraphics[width=0.48\textwidth]{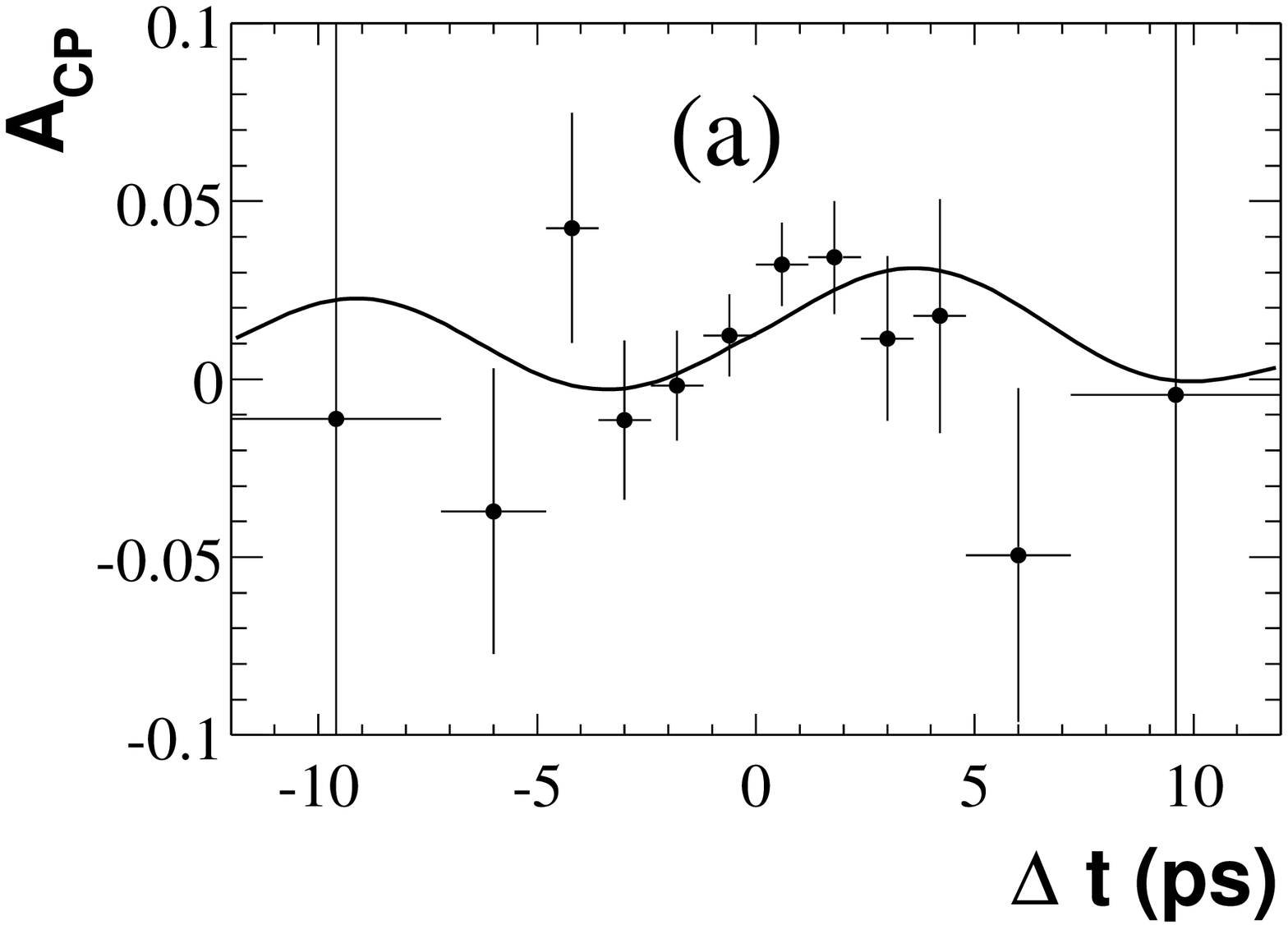}\\
   \includegraphics[width=0.48\textwidth]{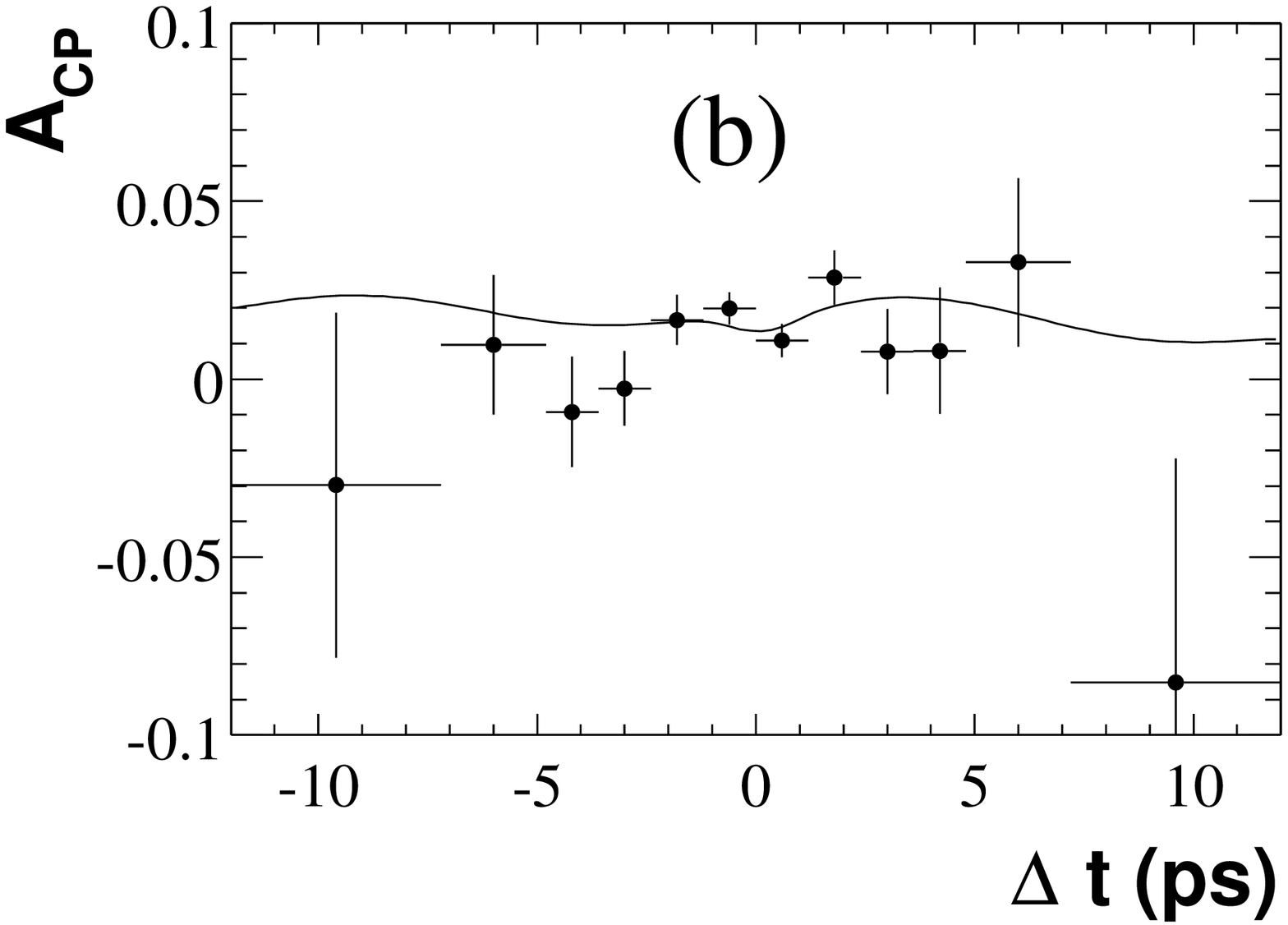}
\end{tabular}
\end{center}
\caption{Raw asymmetry for (a) lepton-tagged and (b) kaon-tagged
  events. The curves represent  the projections of the PDF for the raw
  asymmetry. 
A nonzero value of $a_\dstpi$
would show up as a sinusoidal asymmetry, up to resolution 
and background effects. The offset from the horizontal axis
is due to the nonzero values of $\Delta \epsilon_{\dstpi}$
and $\dmistag{\dstpi}$.}
\label{fig:asym}
\end{figure}

The results of the signal region fit (fourth step) are summarized in
Table~\ref{tab:r1-4}, and the plots of the $\dt$
distributions for the data are shown in
Fig.~\ref{fig:data_sr_Run1-4}
for the lepton-tagged and the kaon-tagged events.
The goodness of the fit has been verified with the Kolmogorov-Smirnov
test and by comparing the likelihood obtained 
in the fit with the likelihood distribution of many parameterized MC experiments
generated with the PDF's obtained in the fit on the data.   
Fig.~\ref{fig:asym} shows the raw, time-dependent \CP asymmetry 
\beq
A(\dt) = {N_{\stag=1}(\dt) - N_{\stag=-1}(\dt) 
		\over N_{\stag=1}(\dt) + N_{\stag=-1}(\dt)}.
\eeq
In the absence of background and with high statistics, perfect tagging, and 
perfect $\dt$ measurement, $A(\dt)$ would be a sinusoidal 
oscillation with amplitude $a_\dstpi$. 
For presentation purposes, the requirements $\mmiss> 1.855$~\gevcc and
$F < 0$ were applied to the data plotted in
Figs.~\ref{fig:data_sr_Run1-4}
and~\ref{fig:asym}, in order to reduce
the background. These requirements were not applied to the fit sample,
so they do not affect our results.

The fitted values of $\Delta m $ reported in Table~\ref{tab:r1-4} 
are in good agreement with the world average $(0.502 \pm 0.007)~\ps^{-1}$ 
\cite{ref:pdg2004}.
The fitted values of the $\Bz$ lifetime need to be corrected for a bias
observed in the simulated samples, $\Delta \tau = \tau_{fit}-\tau_{gen} 
= (-0.03 \pm 0.02)~\ps $  for the lepton-tag and 
$\Delta \tau = (-0.04 \pm 0.02 )~\ps $ for the 
kaon-tag events.
After this correction, the measured lifetimes, $ \tau (\Bz) = 
(1.48 \pm   0.02 \pm 0.02 )~\ps$ and 
$ \tau (\Bz) = (1.49 \pm   0.01 \pm 0.04)~\ps$ 
for the lepton-tag and kaon-tag, 
respectively, are in reasonable agreement with the world average
$ \tau (\Bz) = (1.536 \pm   0.014)~\ps$ \cite{ref:pdg2004}.
The correlation coefficients of $a_\dstpi^\ell$ ($c_\dstpi^\ell$)
with $\Delta m $ and $ \tau (\Bz)$ are  $-0.021$  and  0.019
($-0.060$ and $-0.056$).


\begin{table*}[tbp]
\begin{center}
\caption{\label{tab:r1-4} Results of the fit to the lepton- and kaon-tagged events in the signal region
	$1.845 < \mmiss < 1.880$~\gevcc. 
	Errors are statistical only. See
	Sections~\ref{sec:signalpdf}, \ref{sec:bgd-pdfs}, and~\ref{sec:res} for
	the definitions of the symbols used in this table.}
\begin{ruledtabular}
\begin{tabular}{l|cc|cc}
              & \multicolumn{2}{c|}{Lepton tags} & \multicolumn{2}{c}{Kaon tags}     \\  
\hline
  Parameter description & Parameter & Value                  & Parameter & Value                   \\ 
\hline
Signal weak phase  par.        & $a_{\dstpi}^\ell$ &  $-0.042 \pm 0.019$ & $a_\dstpi^K$  & $-0.025 \pm 0.020$  \\
                     &                   &                     & $b_\dstpi^K$& $-0.004 \pm 0.010$  \\
                     & $c_{\dstpi}^\ell$ &  $-0.019 \pm 0.022$ & $c_\dstpi^K$  & $-0.003 \pm 0.020$   \\
  \hline
Signal $\dt$ PDF & $\dm{\dstpi} $ & $0.518 \pm 0.010$~ps$^{-1}$ & $\dm{\dstpi}$  & $0.4911\pm 0.0076$~ps$^{-1}$          \\
                     & $\lifetime{\dstpi} $ & $1.450 \pm 0.017$~ps  &   $\lifetime{\dstpi}$     & $1.449\pm 0.011$~ps       \\
                     & $\ot{\dstpi}^{\dir}$ &  $0.010 \pm 0.006$    &   $\ot{\dstpi}$           & $0.2302\pm 0.0035$           \\
                     &                      &                       & $\dmistag{\dstpi}$      & $-0.0181\pm 0.0068$          \\
                     & $\Delta \epsilon_{\dstpi}$ &   $0.027 \pm 0.010$  &   $\Delta \epsilon_{\dstpi}$      & $-0.0070\pm 0.0073$    \\
\hline
Signal resolution function & $\bnj{\dstpi}{\cas}$ & $-0.58 \pm 0.16$      &                                   &                 \\
                     & $\bwj{\dstpi}{\cas}$ & $0.23 \pm 2.01$       &                                   &                 \\
                     & $\bnj{\dstpi}{\dir}$ & $0.$ (fixed)          &   $\bn{\dstpi}$           & $-0.255\pm 0.013$            \\
                     & $\bwj{\dstpi}{\dir}$ & $0.$   (fixed)        &   $\bw{\dstpi}$           & $-2.07\pm 0.48$             \\
	             & $\fnj{\dstpi}{\dir}$ & $0.978 \pm 0.008$     &   $\fn{\dstpi}$       & $0.969\pm 0.007$          \\
		     & $\foj{\dstpi}{\dir}$ & $0.$ (fixed)          &   $\fo{\dstpi}$       & $0.000\pm 0.001$          \\
		     & $\snj{\dstpi}{\dir}$ & $1.080 \pm 0.033$     &   $\sn{\dstpi}$          & $1.029\pm 0.023$              \\
		     & $\swj{\dstpi}{\dir}$ & $5.76 \pm 1.44$       &   $\sw{\dstpi}$          & $4.35\pm 0.40$              \\
\hline
Continuum $\dt$ PDF & $\lifetime{\qqbar}$ & 1.26 $\pm$ 0.32~ps     &   $\lifetime{\qqbar}$  & $0.707\pm 0.048 $~ps       \\
		     & $\ot{\qqbar} $ & 0.340 $\pm$ 0.009           &   $\ot{\qqbar}^\tau$        & $0.045 \pm 0.022 $       \\
	      	     & 	        & 	                            &   $\ot{\qqbar}^\delta$          & $0.311\pm 0.006 $            \\
		     & $f^\delta_{\qqbar} $ & 0.815 $\pm$ 0.064     &   $f^\delta_{\qqbar}$         & $0.820 \pm 0.015$          \\
\hline
Continuum resolution function & $\bn{\qqbar}$ & 0.026 $\pm$ 0.048  & $\bn{\qqbar}$           &0.017 $\pm$  0.005    \\
  		     & $\bw{\qqbar}$ & $-0.39 \pm$ 0.23             & $\bw{\qqbar}$           &$-0.043 \pm  0.043$     \\
		     & $\fn{\qqbar}$ & 0.65 $\pm$ 0.12            & $\fn{\qqbar}$           &0.858 $\pm$  0.014     \\
		     & $\fo{\qqbar}$ & 0.068 $\pm$ 0.014            & $\fo{\qqbar}$           &0.018 $\pm$  0.001     \\
		     & $\sn{\qqbar}$ & 0.929 $\pm$ 0.078            & $\sn{\qqbar}$           &1.064 $\pm$  0.008     \\
		     & $\sw{\qqbar}$ & $1.81 \pm 0.28$              & $\sw{\qqbar}$           &2.267 $\pm$  0.099     \\ 
\end{tabular}
\end{ruledtabular}
\end{center}
\end{table*}

\section{SYSTEMATIC STUDIES}
\label{sec:Systematics}

The systematic errors are summarized in Table~\ref{tab:syst}.
Each item below corresponds to the item with the same number in
Table~\ref{tab:syst}.

\begin{itemize}

\item[1.] The statistical errors from the fit in Step~1 are
propagated to the final fit. 
This also includes the systematic errors due to possible differences
between the PDF line shape and the data points.

\item[2.] The statistical errors from the $\mmiss$ 
sideband fit (Step~3) are propagated
to the final fit (Step~4).

\item[3-4.] The statistical errors from the Step~2 
fits are propagated to the final fit.

\item[5.] 
The statistical errors associated with the parameters
obtained from \mc\ are propagated to the final fit.
In addition, the full analysis has been performed on a simulated 
sample to check for a possible bias in the weak phase parameters measured. 
No statistically significant bias has been 
found and the statistical uncertainty of this test has been assigned as a 
systematical error. 

\item[6.] The effect of uncertainties in the 
beam-spot size on the vertex constraint 
is estimated by increasing the beam spot 
size by 50$~\mu$m.

\item[7.] The effect of the uncertainty
in the measured length of the detector in the $z$ direction
is evaluated by applying a 0.6\% variation to the measured values
of $\dt$ and $\dtErr$.

\item[8.] To evaluate the effect of possible misalignments in the
SVT, signal \mc\ events are reconstructed with different
alignment parameters, and the analysis is repeated.

\item[9-11.] The weak phase parameters of the $\btodstrhopm$, peaking, and 
combinatoric
$\BB$ background are fixed to 0 in the fits. To study the effect of
possible interference between $\btou$ and $\btoc$ amplitudes
 in these backgrounds, their weak phase parameters are
varied in the range  $\pm 0.04$ and the Step-4 fit is repeated.
We take the largest variation in each weak phase parameter as its systematic error.

\item[12.] In the final fit, we take the values of the
parameters of $\T'_\peak$ from a fit to simulated peaking
\BB\ background events.
The uncertainty due to this
is evaluated by fitting the simulated sample, setting the 
parameters of $\T'_\peak$ to be identical to those of $\T'_\comb$.

\item[13.] The uncertainty due to possible differences between the
$\dt$ distributions for the combinatoric background in the $\mmiss$
sideband and signal region is evaluated by comparing the results of
fitting the simulated sample with the $\T'_\comb$ parameters taken
from the sideband or the signal region.

\item[14.] The ratio $f_\dstrho/f_\dstpi$ is varied by 
the uncertainty in the corresponding ratio of branching fractions,
obtained from Ref.~\cite{ref:pdg2004}.

\end{itemize}

\begin{table*}
\begin{center}
\caption{\label{tab:syst} Systematic errors in $a_\dstpi^\ell$ and $c_\dstpi^\ell$ for 
	lepton-tagged events and $a_\dstpi^K$, $b_\dstpi^K$, and $c_\dstpi^K$ for 
	kaon-tagged events. }
\begin{tabular}{|l|cc|ccc|}  
\hline\hline
Source     & \multicolumn{5}{c|}{Error $(\times 10^{-2})$ } \\
\cline{2-6} 
           & \multicolumn{2}{c|}{Lepton tags} & \multicolumn{3}{c|}{Kaon tags} \\
\cline{2-6}
		&  $ a_\dstpi^\ell$   
		&  $ c_\dstpi^\ell$ 
	        & $a_\dstpi^K$   
		&  $b_\dstpi^K$  
		& $c_\dstpi^K$  
   \\ 
\hline
1. Step~1 fit 			& 0.04 & 0.04    & $0.10 $& $0.04 $       & $0.04 $         \\
2. Sideband statistics  	& 0.08 & 0.08   &  $0.40  $   &   $0.12  $  &  $0.44  $   \\
3. $\fmis_\dstpi$        	& 0.02 & 0.02   &  $0.02  $   &   negl.  $  $  &  negl.    \\
4. $\rho_\dstpi$        	& 0.02 & 0.02   &  $0.02  $   &   negl.   &  negl.   \\
\hline
5. MC statistics    		& 0.60  & 0.82   &  $0.68  $   &   $0.34  $  &  $0.70  $   \\ 
\hline
6. Beam spot size      		& 0.10	& 0.10 &  $0.07  $   &   $0.13  $  &  $0.06  $   \\
7. Detector $z$ scale       	& 0.03	& 0.03 &  $0.02  $   &   negl.               &  $0.03  $   \\
8. Detector alignment 		& 0.25   & 0.55   & $0.25  $ & $0.13  $    & $0.41  $   \\
\hline
9. Combinatoric background weak phase par.     & 0.25  & 0.22   & $0.80$  & $0.56  $   &  $0.72  $   \\
10. Peaking background weak phase par.   & 0.36  & 0.38 &  $0.29$   &  $0.17  $   &  $0.27  $   \\
11. $D^*\rho$ weak phase par.    & 0.53  & 0.52   &  $0.57  $   &  $0.58  $   &  $0.58  $   \\ 
\hline
12. Peaking background   & 0.21    & 0.31   &  $0.21  $   &   $0.41  $  &  $0.31  $   \\
13. Signal region/sideband difference     & negl.    & negl. & $0.04$  &   $0.03$  &  $0.05$   \\
14. \BR($\btodstrhopm$)  &  0.17   & 0.33   &  $0.17  $   &   $0.22  $  &  $0.33  $   \\
\hline
Total systematic error           & 1.0    & 1.3   &  $1.4  $   &  $1.0  $   &  $1.5 $\\ 
\hline
Statistical uncertainty     & 1.9 & 2.2 & 2.0 & 1.0 & 2.0 \\
\hline\hline
\end{tabular}
\end{center}
\end{table*}

\section{PHYSICS RESULTS}
\label{sec:Physics}

Summarizing the values and uncertainties of the weak phase parameters, we obtain
the following results from the lepton-tagged sample:
\bea
a_\dstpi^\ell   &=& 
	-0.042 \pm 0.019 \pm 0.010, \nonumber\\
c_\dstpi^\ell   &=& 
	-0.019 \pm 0.022 \pm 0.013. 
\label{eq:S-lepton}
\eea
The results from the kaon-tagged sample fits are
\begin{eqnarray}
a_\dstpi^K   &=& 
	-0.025 \pm 0.020 \pm 0.013, \nonumber\\
b_\dstpi^K   &=& 
	-0.004 \pm 0.010 \pm 0.010, \nonumber\\
c_\dstpi^K &=& 
	-0.003 \pm 0.020 \pm 0.015.
\label{eq:abc-kaon}
\eea
Combining the results for lepton and kaon tags 
gives the amplitude of the time-dependent \CP asymmetry,
\begin{eqnarray}
a_\dstpi &=& 2\r\sin(2\beta+\gamma)\cos\deltaPhase  \nonumber \\
&=&	-0.034 \pm 0.014 \pm 0.009,
\label{eq:combined-a}
\end{eqnarray}
where the first error is statistical and the second is systematic.
The systematic error takes into account correlations between
the results of the lepton- and kaon-tagged samples coming from 
the systematic uncertainties related to detector effects,
to interference between $\btou$ and $\btoc$ amplitudes
 in the backgrounds and from \BR($\btodstrhopm$).
This value of $a_\dstpi$ deviates from zero by 2.0 standard deviations.

Previous results of time-dependent \CP asymmetries related 
to $2\beta+\gamma$ appear in 
Ref.~\cite{ref:run1-2,ref:others}. This measurement supersedes the results
of the partial reconstruction analysis reported in 
Ref.~\cite{ref:run1-2} and improves the precision on 
$a_\dstpi$ and $c_\dstpi$ with respect to the
average of the published results.

We use a frequentist method, inspired by Ref.~\cite{ref:Feldman}, to 
set a constraint on $2\beta+\gamma$. To do this, we need a value
for the ratio $\r$ of the two interfering amplitudes.
This is done with two different approaches.

In the first approach, to avoid any assumptions on the value 
of $\r$,  
we obtain the lower limit on $|\sin(2\beta+\gamma)|$
as a function of $\r$.

We define a $\chi^2$
function that depends on $\r$, $2\beta+\gamma$, and $\deltaPhase$:
\beq
\chi^2(\r,2\beta+\gamma , \deltaPhase) = \sum_{j,k=1}^3 \Delta x_j 
	V^{-1}_{jk} \Delta x_k, 
\label{eq:chi2}
\eeq
where 
$\Delta x_j$ is the difference between the result of our measurement
of $a_\dstpi^K$, $a_\dstpi^\ell$, or $c_\dstpi^\ell$ (Eqs.~(\ref{eq:abc-kaon}) 
and~(\ref{eq:S-lepton})) and the
corresponding theoretical expressions given by Eq.~(\ref{eq:abc}).
We fix $\r$ to a trial value $r^0$.
The measurements of $b_\dstpi^K$ and $c_\dstpi^K$ are not used in the fit,
since they depend on the unknown values of $r'$ and $\delta'$.
The measurement error matrix $V$ is nearly diagonal, and
accounts for correlations between the measurements due to correlated
statistical and systematic uncertainties.
%
We minimize $\chi^2$ as a function of $2\beta+\gamma$ and
$\deltaPhase$, and obtain $\chi^2_{min}$, the minimum value of
$\chi^2$. 

In order to compute the confidence level for a given value $x$ of
$2\beta+\gamma$, we perform the following procedure:
\begin{enumerate}
\item We fix the value of $2\beta+\gamma$ to $x$ and minimize $\chi^2$
as a function of $\deltaPhase$.
We define $\chi'^2_{min}(x)$ to be the minimum value
of the $\chi^2$ in this fit, and $\deltaPhase_{toy}$ to be the fitted
value of $\deltaPhase$.
We define $\Delta\chi^2(x) \equiv
\chi'^2_{min}(x) - \chi^2_{min}$.

\item We generate many parameterized MC experiments with the same
sensitivity as the data sample, taking into account correlations
between the observables, expressed in the error matrix $V$ of
Eq.~(\ref{eq:chi2}). To generate the observables $a_\dstpi^K$,
$a_\dstpi^\ell$, and $c_\dstpi^\ell$, we use the values
$(2\beta+\gamma)=x$, $\r=r^0$ and
$\deltaPhase=\deltaPhase_{toy}$. For each experiment we calculate the
value of $\Delta\chi^2(x)$, computed with the same procedure used for
the experimental data.

\item We interpret the fraction of these experiments for which
$\Delta\chi^2(x)$ is smaller than $\Delta\chi^2(x)$ in the data to be
the confidence level (CL) of the lower limit on
$(2\beta+\gamma)=x$.
\end{enumerate}

The resulting 90\% CL lower limit on $|\sin(2\beta+\gamma)|$ as a function of 
$\r$ is shown in Fig.~\ref{fig:limit-vs-r}. The $\chi^2$ function is invariant under the 
transformation $ 2\beta+\gamma \to \pi/2 + \delta^*$ and  $ \delta^* \to \pi/2 - 2\beta+\gamma $. The limit shown in Fig.~\ref{fig:limit-vs-r} is always the weaker of 
these two possibilities.

\begin{figure}[htb]
\begin{center}
        \includegraphics[width=0.48\textwidth]{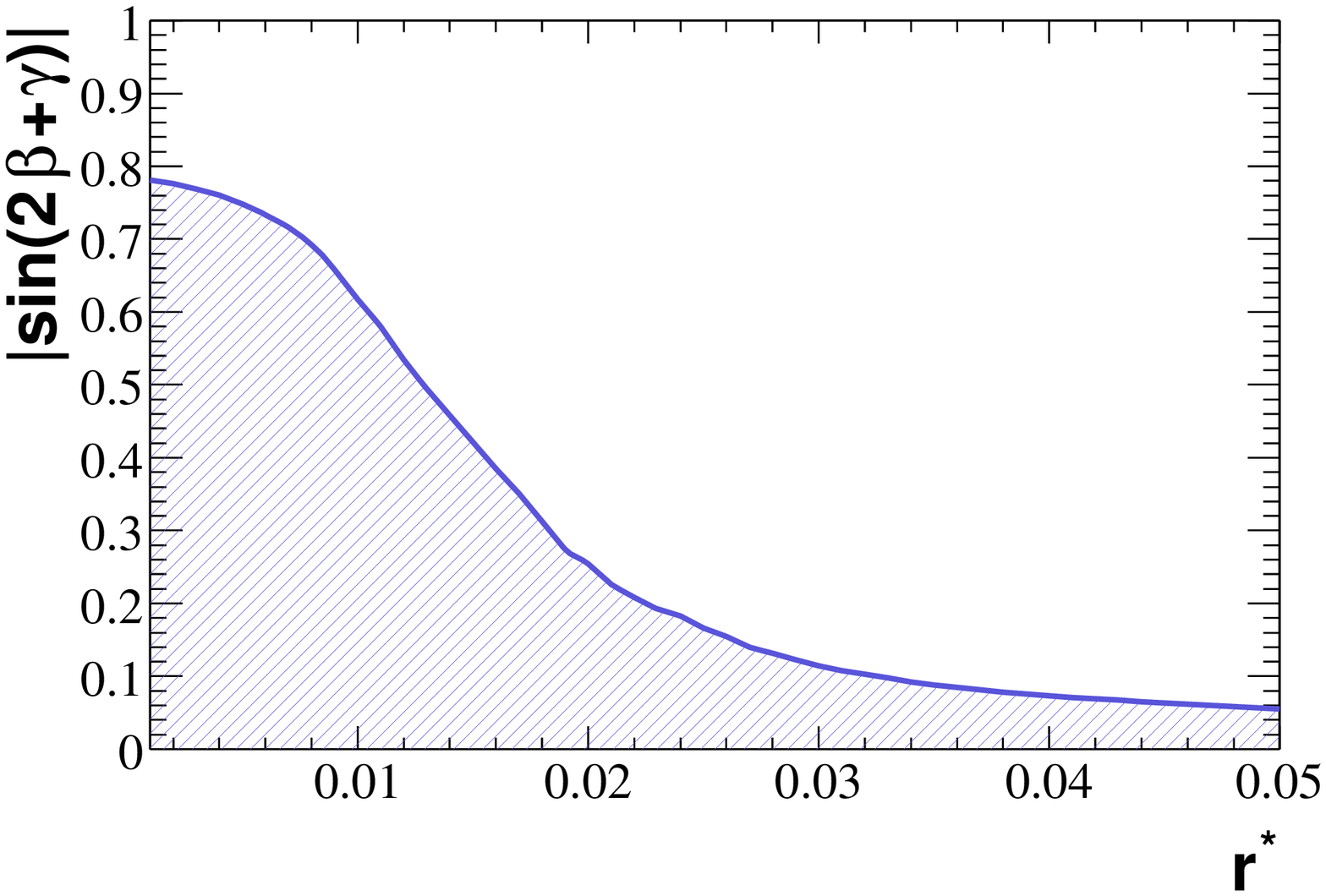}
\end{center}
\vspace*{-0.5cm}
\caption{
 Lower limit on $|\sinphi|$ at 90\% CL as a function of $\r$,
for $\r>0.001$.
}
\label{fig:limit-vs-r}
\end{figure}

In the second approach, we estimate $\r$  as originally proposed in 
Ref.~\cite{ref:book}, and assume SU(3) flavor symmetry.
With this assumption, $\r$ can be estimated from the
Cabibbo angle $\theta_C$, the ratio of branching fractions ${\cal
B}(B^0\rightarrow {\Dstar}_s^{+} \pi^-) / {\cal B}(B^0\rightarrow
{\Dstar}^{-} \pi^+)=(5.4^{+3.4}_{-3.7} \pm 0.7)\times 10^{-3}$~\cite{ref:Dspi}, 
and the ratio of decay constants
$f_{\Dstar_s} / f_{\Dstar}=1.10 \pm 0.02$~\cite{ref:dec-const}, 

\begin{eqnarray}
\r &=& \sqrt{\frac{{\cal B}(B^0\rightarrow {\Dstar}_s^{+} \pi^-)} 
{{\cal B}(B^0\rightarrow
{\Dstar}^{-} \pi^+) } }
\,  \frac{f_{\Dstar}} {f_{\Dstar_s}} \, \tan (\theta_C),
\label{eq:rformula}
\end{eqnarray}
yielding the measured value
\beq
\rmeas = 0.015^{+0.004}_{-0.006}.
\label{eq:rmeas}
\eeq
This value depends on the value of ${\cal B}(D_s^+\to \phi\pi^+)$,
for which we use our recent measurement~\cite{ref:phipi}.

Equation~(\ref{eq:rformula}) has been obtained with two approximations. 
In the first approximation, the exchange diagram amplitude $E$ contributing to the decay
$B^0\rightarrow {\Dstar}^{+} \pi^-$ has been neglected and only the 
tree-diagram amplitude $T$ has been considered. 
Unfortunately, no reliable estimate of the
exchange term for these decays exists.
The only decay mediated by an exchange diagram for which the rate 
has been measured is the Cabibbo-allowed decay $ \Bz \to D_s^{-}
K^{+}$. The average of the \babar\ and Belle 
branching fraction measurements~\cite{ref:Dspi,ref:belledsk}  
is $(3.8 \pm 1.0)\times 10^{-5}$. 
This yields the approximate ratio
$ {\cal B}(\Bz \to D_s^{-} K^{+}) / {\cal B}(\Bz \to D^{-} \pi^{+}) \sim 10^{-2}$,
which confirms that the exchange
diagrams are strongly suppressed with respect to the tree diagrams.
Detailed analyses~\cite{ref:topolo} of the $B \to D \pi$ and $B \to D^* \pi$ decays
in terms of the topological amplitudes  conclude
that $|E'/T'| = 0.12 \pm 0.02$ 
for $\Bz \to D^- \pi^+$ and $|\bar{E}/\bar{T}| < 0.10$ for $\Bz \to {\Dstar}^{-} \pi^+$ decays, where $E'$, $\bar{E}$ and $T'$, $\bar{T}$ are the exchange and 
tree amplitudes for these Cabibbo-allowed decays.
We assume that a similar suppression holds for the Cabibbo-suppressed 
decays considered here.

The second approximation involves the use of the ratio 
of decay constants
$f_{\Dstar} / f_{\Dstar_s}$ to take into account
SU(3) breaking effects and assumes factorization. 
We attribute
a 30\% relative error to the theoretical assumptions
involved in obtaining the value of $\r$ of Eq.~(\ref{eq:rmeas}), and use it as described below.

We add to the $\chi^2$ of Eq.~(\ref{eq:chi2}) the term
$\Delta^2(\r)$ that
takes into account both the
Gaussian experimental errors of Eq.~(\ref{eq:rmeas}) 
and the 30\% theoretical uncertainty according to the prescription 
of Ref.~\cite{Hocker:2001xe}: 
\beq
\Delta^2(\r) = \left\{\begin{matrix} 
	\left(\dfrac{\r - 1.3\, \rmeas }{ 0.004}\right)^2  & , &       
		\xi_\r          > 0.3 &, \\[5mm]
	0 & , & \left|\xi_\r\right| \le 0.3 &,\\[2mm]
	\left(\dfrac{\r - 0.7\, \rmeas }{ 0.006}\right)^2 & , &       
		\xi_\r          < -0.3 &,
\end{matrix}\right.  
\eeq
where $\xi_\r \equiv {(\r - \rmeas) / \rmeas}$. 

To obtain the confidence level we have repeated the procedure described
above with the following changes. To compute $\chi^2_{min}$ 
we minimize $\chi^2$ as a function of $2\beta+\gamma$, $\r$ and
$\deltaPhase$. The value $\chi'^2_{min}(x)$ is obtained minimizing
$\chi^2$  as a function of $\r$ and
$\deltaPhase$, having fixed $2\beta+\gamma$ to a given value $x$.
We define $\deltaPhase_{toy}$  and $\r_{toy}$ to be the fitted
value of $\deltaPhase$ and $\r$ in this fit. 
To generate the observables $a_\dstpi^K$,
$a_\dstpi^\ell$, and $c_\dstpi^\ell$ in the parameterized MC experiments, 
we use the values
$(2\beta+\gamma)=x$, $\r=\r_{toy}$ and
$\deltaPhase=\deltaPhase_{toy}$.

The confidence level as a function of $|\sin(2\beta+\gamma)|$ is 
shown in Fig.~\ref{fig:FC1}. We set the lower limits 
$|\sinphi|> 0.62~(0.35)$ at 68\% (90\%) CL.
The implied probability contours for the apex of the unitarity triangle,
parameterized in terms of $\bar \rho $ and $\bar \eta$ defined 
in Ref.~\cite{ref:wolfen}, 
appear in Fig.~\ref{fig:UT}.

\begin{figure}[tb]
\begin{center}
        \includegraphics[width=0.48\textwidth]{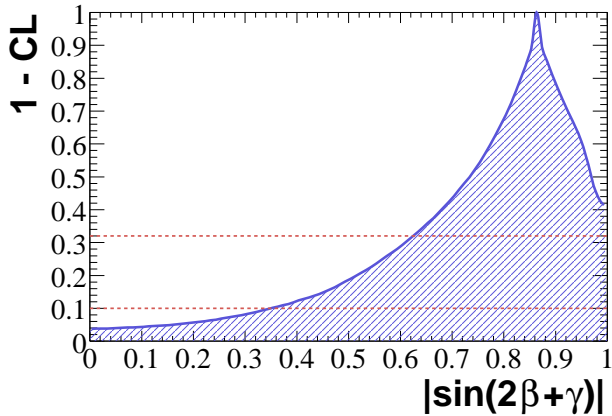}
\end{center}
\vspace*{-0.5cm}
\caption{The shaded region denotes the allowed range of $|\sinphi|$
for each confidence level. The horizontal lines show, from top
to bottom, the 68\% and 90\% CL. 
}
\label{fig:FC1}
\end{figure}

\begin{figure}[htb]
\begin{center}
        \includegraphics[width=0.48\textwidth]{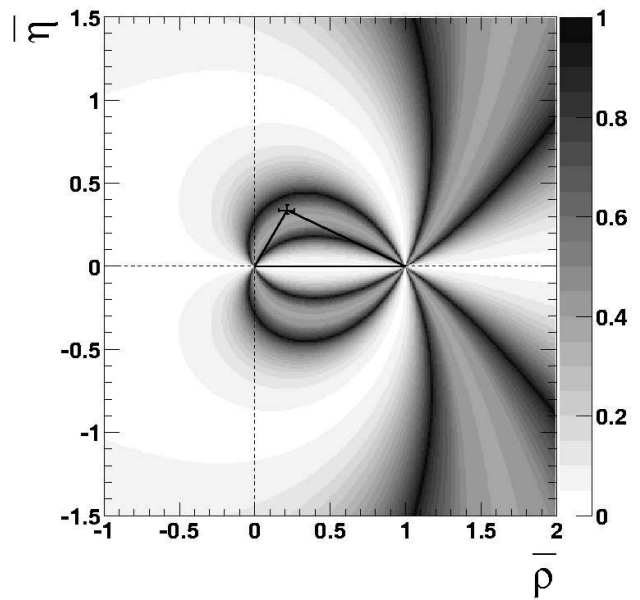}
\end{center}
\vspace*{-0.5cm}
\caption{
Contours of constant probability (color-coded in percent)
for the position of the apex of the unitary triangle
to be inside the contour, based on the results of Fig.~\ref{fig:FC1}.
The cross represents the value
and errors on the position 
of the apex of the unitarity triangle from the 
CKMFitter fit using the ``ICHEP04'' results excluding 
this measurement \cite{Hocker:2004}.
}
\label{fig:UT}
\end{figure}

\section{SUMMARY}
\label{sec:Summary}

We present a  measurement of the time-dependent \CP asymmetries 
 in a sample of partially reconstructed
$\btodstpi$ events.  In particular, we have measured
 the parameters related to $2\beta+\gamma$ to be 
\begin{eqnarray}
a_\dstpi &=& 2\r\sin(2\beta+\gamma)\cos\deltaPhase  \nonumber \\
&=&	-0.034 \pm 0.014 \pm 0.009
\label{eq:combined-sum}
\end{eqnarray}
and 
\bea
c_\dstpi^\ell   &=& 2 \r \cos(2\beta+\gamma)\sin\deltaPhase \nonumber \\
   &=& 	-0.019 \pm 0.022 \pm 0.013, 
\label{eq:summary-c}
\eea
where the first error is statistical and the second is systematic. 
We extract limits as a function of the ratio $\r$
of the $b\rightarrow u \overline c
d$ and $b\rightarrow c \overline u d$ decay amplitudes.
With some theoretical assumptions, we
interpret our results in terms of the lower limits
$|\sinphi|> 0.62~(0.35)$ at 68\% (90\%) CL.

\section{Acknowledgments}
\label{sec:Acknowledgments}

We are grateful for the 
extraordinary contributions of our \pep2\ colleagues in
achieving the excellent luminosity and machine conditions
that have made this work possible.
The success of this project also relies critically on the 
expertise and dedication of the computing organizations that 
support \babar.
The collaborating institutions wish to thank 
SLAC for its support and the kind hospitality extended to them. 
This work is supported by the
US Department of Energy
and National Science Foundation, the
Natural Sciences and Engineering Research Council (Canada),
Institute of High Energy Physics (China), the
Commissariat \`a l'Energie Atomique and
Institut National de Physique Nucl\'eaire et de Physique des Particules
(France), the
Bundesministerium f\"ur Bildung und Forschung and
Deutsche Forschungsgemeinschaft
(Germany), the
Istituto Nazionale di Fisica Nucleare (Italy),
the Foundation for Fundamental Research on Matter (The Netherlands),
the Research Council of Norway, the
Ministry of Science and Technology of the Russian Federation, and the
Particle Physics and Astronomy Research Council (United Kingdom). 
Individuals have received support from 
CONACyT (Mexico),
the A. P. Sloan Foundation, 
the Research Corporation,
and the Alexander von Humboldt Foundation.

\end{document}